\documentclass[reprint, twocolumn,floatfix]{revtex4}
\usepackage{hyperref}
\usepackage[english]{babel}
\usepackage[latin1]{inputenc}
\usepackage{enumerate}
\usepackage{amsfonts}
\usepackage{amssymb}
\usepackage{amsmath}
\usepackage{dcolumn}
\usepackage{bm}
\usepackage{graphicx, graphics}
\usepackage{graphicx}
\usepackage{xcolor}
	\begin{document}
		\title{Drag Force in the Vacuum of Confining Gauge Theories}

\author{Saulo Diles}
\email{smdiles@ufpa.br}
\affiliation{Campus Salin\'opolis,\\ Universidade Federal do Par\'a,\\
68721-000, Salin\'opolis, Par\'a, Brazil}

\author{Miguel Angel Martin Contreras}
\email{miguelangel.martin@uv.cl}
\affiliation{%
 Instituto de F\'isica y Astronom\'ia, \\
 Universidad de Valpara\'iso,\\
 A. Gran Breta\~na 1111, Valpara\'iso, Chile
}

\author{Alfredo Vega}%
 \email{alfredo.vega@uv.cl}
\affiliation{%
 Instituto de F\'isica y Astronom\'ia, \\
 Universidad de Valpara\'iso,\\
 A. Gran Breta\~na 1111, Valpara\'iso, Chile
}
\begin{abstract}
The complete absence of isolated quarks reaching particle detectors after high energy collisions suggests that some physical mechanism generates resistance to their propagation in the vacuum. In order to reveal such a mechanism, we analyze the fate of an infinitely heavy quark that is initially propagating in the vacuum with inertial motion. The non-perturbative structure of the vacuum is treated here using the gauge/ gravity correspondence, the isolated quark on the boundary gauge theory is dual to a trailing string moving in the bulk of a higher dimensional curved space. We find that, for a large class of non-conformal gauge theories with a holographic dual, the geometrical structure of the bulk geometry induces a drag force on the quark that moves in the vacuum. In addition, we show that for these gauge theories there will be the presence of such a drag force due to its vacuum whenever the dual bulk geometry generates a linear potential for a $q\bar{q}$ pair. The relation of the linear $q\bar{q}$ potential with the drag force on the isolated quark is a holographic piece of evidence that both phenomena are different manifestations of the confinement of quarks.

\end{abstract}

\maketitle
\section{\label{sec:level1}Introduction}
In the present paper we discuss the confinement of quarks using holography. We treat the confinement problem from the point of view of a single  \emph{free} quark with infinite mass, considered as a probe particle. The physical picture is set by analyzing the relativistic quark's motion. This massive probe quark is dual to an open string stretching inside the bulk of the 5-dimensional AdS space, with one of its ends tied to the conformal boundary. The other end is considered moving freely since the scenario is not thermal, thus there is no associated black hole event horizon to tie up this end. This implies that at the initial stage this free end is only affected by the AdS background. This momentum transfer between the background and the string defines the drag experienced by the probe.  
 
 The field theory describing strong interactions, the so called \emph{Quantum Chromodynamics} (QCD)
 has two very important features:
asymptotic freedom  \cite{Gross:1992cw} and confinement. The asymptotic freedom  was very important for establishing QCD as the formal theory of strong interactions. For the high energy phenomena, this theory is perturbative and leads to a wide range of important results \cite{Dokshitzer:1991wu,Brock:1993sz}.  On the other hand, confinement is harder to deal   with its complexity \cite{Greensite:2003bk, Alkofer:2006fu}, and also because QCD is strongly coupled field theory at the  low energy limit, where the perturbative treatment is not possible. This is translated into a non-perturbative vacuum.  

There are two definitions of confinement, quark confinement and color confinement. The former establishes that no single quark can be measured as an asymptotic state in any particle detector. Color confinement means that no color charged (or colored) particle state can be measured as an asymptotic state in any particle detector. It happens that color confinement includes quark confinement since quarks are colored particles. 

A great difficulty understanding the physics of confinement comes when the strongly coupled interactions at low energy are considered, since confinement is manifest on the observables related with low energy physics, such as the hadronic Regge trajectories  and the slope of the $q\bar{q}$ potential at very large separations, implying that theoretical description of confinement is intrinsically  non-perturbative.

The AdS/CFT correspondence \cite{Aharony:1999ti,Witten:1998qj} establishes a connection between the non-perturbative regime of a gauge theory and a gravity theory at weak coupling.  The holographic picture of confinement was established by the dependence of the interaction potential of a static quark--anti-quark pair ($q\bar{q}$)   with the separation distance between them. The associated gauge theory will be confining if the potential is linear for large separation. In the paper by Kinar \emph{et. al.} \cite{Kinar:1998vq}  was established a direct mathematical criterion to determine whether the gauge theory is confining or not by finding a sufficient condition for the $q\bar{q}$ potential to be linear
at large separations. In this sense, the physical  picture of confinement is clear for the  static $q\bar{q}$ pair system. 

In order to complete the physical picture of confinement we analyse the case of a single quark that is initially moving \emph{freely} (i.e. not bounded) in the vacuum of the  gauge theory. To do so, we extend the  Gubser's proposal \cite{Gubser:2006bz} to the case of zero temperature by providing the dual picture of a free quark on the
gauge theory vacuum by an open (trailing) string in the bulk with one endpoint attached on the quark location  at the boundary and the other end moves untied. By analyzing such a dual string configuration, we find that for a large class of gauge theories with a holographic dual,  there will be a drag force acting on the quark moving in the vacuum if the such theories are confining.  Another analogy comes from electricity: conformal vacuum, where the lack of drag force is translated into no confinement, is similar to a \emph{electric conductor}, where carriers are free to move. On the other case, not conformal vacuum, where drag force implies confinement, is equivalent to a \emph{dielectric}, characterized by  carriers that cannot move freely.  In this simple analogy, confinement and electrical resistant are \emph{emergent properties} arising from the electrical (conformal) properties of the medium. The proposed holographic analysis discussed here provides this nice physical picture for the vacuum of gauge theories experienced by the isolated probe quark. \textcolor{black}{ Another possible backgrounds that have been considered in the literature are the finite temperature case \cite{Talavera:2006tj}, finite chemical potential \cite{Cheng:2014fza}, and the strong magnetic field \cite{Mamo:2016xco}}.

The presence of a drag force due to the vacuum looks like to be an effect associated with the averaged quantum interactions of the probe quark with the vacuum fluctuations. Effectively the quark motion is damped or equivalently the quark exerts some work on the medium so that its kinetic energy is transferred to the medium. On a phenomenological point of view, when an isolated quark is scattered in the vacuum with sufficiently high energy there will be formed a jet of hadrons centred in the initial direction of the moving quark. This  is the jet formation phenomenon \cite{Ali:2010tw}. Our hypothesis is that this drag force due to the vacuum is directly related with the jet formation process. In the  holographic picture, the long string attached to the quark feels a gravitational inertia (resistance) on the bulk  breaking it down into many
pieces, each with both ends attached to the boundary in a semi-classical transition, i.e., the free quark interacting with the gauge theory vacuum \emph{hadronizes}. Here we use the results for drag force on the vacuum to estimate the jet energy as a function of the initial quark energy.


The work is organized as follows: in section \ref{quark-vacuum}, we do a summary of the quark-vacuum interaction in the perspective of the fully inclusive quark-jet correlation function and its interpretation in terms of the energy lost by a parton moving in the vacuum. In section \ref{section-SYM} we discuss the drag force in the vacuum of a $\mathcal{N}=4$ SYM at zero temperature, probing that there is no drag force in such a model. In section \ref{section-3}, we discuss the bottom-up of wall model at zero temperature case, demonstrating the existence of drag force attached to the presence of the energy scale that breaks softly the conformal invariance. In section \ref{section-4}, we discuss the general case given by an AdS-like background. The existence of a drag-force in such backgrounds is conditioned by the existence of a geometric upper bound in the 4-momentum transferred by the string from the boundary to the bulk. In sections \ref{section-5} and \ref{section-6} we focus on the top-down models D3/D7 brane system and the Sakai-Sugimoto configuration. As in the bottom-up case, the drag force is conditioned again by the existence of a confining  scales in both models. In section \ref{section-7} we estimated the fraction of energy loss, finding that the non-conformal systems have associated the so called \emph{dead cone effect}. Finally, in section \ref{section-8}, we give some conclusions and final remarks about the phenomenology discussed here. 

\section{Quark-Vacuum interaction phenomenology}\label{quark-vacuum}

One of QCD's essential properties is confinement, implying that no single-colored parton can exist as a stable free particle. After high energetic collisions, such deep inelastic scattering \cite{Chen:2006vd},  free partons travel until they hadronize by mechanisms as the gluon bremsstrahlung \cite{Webber:1999ui}. Recall that partons are not measured since they do not reach the detector due to their short mean free path compared to the detector's distance. In other words, these free partons produce jets of hadrons that will be detected short after. Therefore, talking about hadronization is tightly connected to the partonic propagation mechanism.  However, this transition from free partons to bounded hadrons and the jet formation have not been deeply explored. See for example \cite{Accardi:2019luo}. 

One of the forms to address the interaction between free partons and the hadronic jets is the quark-jet correlator, which accounts for the hadron formation in the physical region $x<1$, where $x$ is the Bjorken variable \cite{Accardi:2017pmi,Collins:2007ph,Sterman:1986aj} 

\begin{multline}
\Xi_{ij}(k,\hat{n})=\frac{1}{N_c}\int{d^4\,\eta\,e^{i\,\,k\cdot \eta}\text{Tr}_c\,\langle 0\left|\mathcal{T}\,W_1(\infty,\eta,
\hat{n})\,\psi_i(\eta)\,\right.}\\
\left.
\times \bar{\mathcal{T}}\bar{\psi}_j(0)\,W_2(0,\infty,\hat{n})\right| 0\rangle   
\end{multline}

\noindent where we have the following definitions: $k$ is the quark four-momentum, $\psi(\eta)$ is the spinor field associated to the quark,  $\mathcal{T}$ stands for time ordering whereas $\bar{\mathcal{T}}$ is the anti-time ordering and $\left| 0\rangle \right.$ is the non-perturbative QCD vacuum. Flavor indices are ommited for simplicity. The gauge invariance structure is guaranteed by the Wilson line operators $W(\infty,\hat{n},\eta)$, conditioned by choice of the line directions given by the unitary vector $\hat{n}$. The choice of the Wilsonian path is necessary in order to apply QCD factorization theorems. Finally, the vector $\hat{n}$ parametrizes the formed jet direction. The trace over color is required since the jets should be colorless when they reach the detectors. This correlator is also called \emph{fully inclusive} since, from the pure diagrammatically point of view,  it describes all the  hadronic products appearing due to the quark-vacuum interaction without considering jet reconstruction. 

This correlator can be parametrized in terms of jet-parton correlation functions using a Lorentz invariant Dirac decomposition, depending on the final hadron twist.  For example, in the case of twist-3 order, the jet-quark correlator takes the following spectral form: 

\begin{equation}
\Xi(k^2)=\int{d\,\sigma^2\,\left[J_1(\sigma^2)\,\sigma\,\textbf{1}+J_2(\sigma^2)\gamma_\mu\,k^\mu\right]\,\delta(k^2-\sigma^2)}    
\end{equation}

\noindent with $\sigma^2$ defined as the invariant mass of the outgoing jet; the functions $J_i(\sigma^2)$ are the jet mass distributions satisfying

\begin{equation}
J_2(\sigma^2)\geq J_1(\sigma^2)\geq0\,\, \text{and}\,\int{d\,\sigma^2\,J_2(\sigma^2)=1}.     
\end{equation}

\noindent imposed by CPT invariance. Notice that $\sigma^2$ can be directly connected, by energy conservation, with the energy lost by a parton moving in the QCD vacuum. The drag force exerted by the vacuum is translated into the quark fragmentation that leads to the production of jets.

Jet formation is the observed phenomenon happening when an isolated quark is scattered away from its bounded partners. This phenomenology is assigned to the interaction of the colored quark with the QCD vacuum. The physical mechanism responsible for transmuting the initial quark in a jet of hadrons has been investigated for a long time \cite{Casher:1974vf, Field:1977fa, Takagi:1979wn}.  The non-perturbative nature of QCD vacuum makes it a challenger and exciting topic which motivated different models such as stochastic instanton configurations \cite{Simonov:1987rn, Shuryak:1992jz}, chiral disorder \cite{Janik:1998ki} and dual superconductors \cite{Nair:1984rg, Hosek:1993bw, DiGiacomo:1999yas, Agasian:1999id}. However, a complete description of the non-perturbative QCD vacuum is still lacking. 

It is interesting to note that the QCD vacuum also plays a role in a dense media. The description of the energy loss by gluon radiation of a quark moving in a dense media can be written as an opacity expansion whose zero-order term persists even for zero matter density. It is attributed to the QCD vacuum itself \cite{Armesto:2003jh,Zhang:2007ai}. Even that the opacity expansion assumes the presence of a medium, it reveals the vacuum itself acts on the propagating quark damping its motion. It is also remarkable that the vacuum contribution to the energy lost by the heavy quark reproduces the dead cone effect.

In the context of the gauge/ gravity correspondence, the description of quark-vacuum interaction  has focused on colorless hadronic systems, especially in the description of mesons. The present paper introduces the analysis of quark-vacuum interaction by extending the Gubser proposal \cite{Gubser:2006bz} to a vacuum medium. The concept of vacuum in this approach comes from the Minkowski signature of the metric and the absence of a black role in the bulk geometry, which means that the boundary gauge theory is at zero temperature and zero chemical potential. As a result, as we will expose in Section \ref{section-4},  a non-vanishing drag force is allowed, and consequently, will be there to act on a single quark propagating in the vacuum of the confining gauge theory. 

 \section{The vacuum of $\mathcal{N}=4$ SYM}\label{section-SYM}
  In this section we analyze  the vacuum of the conformal dual field theory $\mathcal{N}=4$ SYM, where there is no confinement \cite{Rey:1998ik, Maldacena:1998im}. This conformal theory is constructed as the dual of the geometric background generated by $N_c$ coincident D3-branes, that is AdS$_5\times\,S^5$. See  \cite{Gubser:2006bz}.

 Let us focus on the AdS part of such geometry, parametrized by the
 5-dimensional  Poincar\'e patch in the following form:
 
 \begin{equation}
     dS^2 = \frac{R^2}{z^2}\left[-dt^2+d\vec{x}^2 +dz^2\right],\,z\in(0,\infty).
 \end{equation}

The isolated  heavy quark is dual to an end of an open (trailing) string, attached to the boundary and hanging into the AdS bulk. The other string end just goes to IR sector of AdS$_5$.


The probe quark is assumed to move at constant velocity. Also we adopt the same ansatz used in the finite temperature case for the  string profile: $X^\mu(t,z) = (t, x(t,z),0,0,z),~~x(t,z) = vt+\xi(z),$ where $v$ is the quark velocity (in the lab frame). For this string  configuration the Nambu-Goto action is given by

\begin{equation}
    \mathcal{S} =  \frac{R^2}{2\pi \alpha'}\int dt\,dz\, \frac{1}{z^2}\,\sqrt{1-v^2 + \xi'^2}.
\end{equation}

When we use this ansatz, the  Lagrangian density $\mathcal{L} =  \frac{1}{2\pi \alpha'z^2}\sqrt{1-v^2 + \xi'^2}$ becomes a function of $z$ and $\xi'(z)$ only. Therefore, the equation of motion for the string  is expressed as the conservation of
$\pi_\xi = \frac{\partial\mathcal{L}}{\partial \xi'}$, that is given by

\begin{equation}
\pi_\xi = \frac{R^2}{2\pi\alpha'}\frac{\xi'}{z^2\sqrt{1-v^2 + \xi'^2}}.
\end{equation}

From the expression above, we can solve for $\xi(z)$, obtaining 

\begin{equation}
\frac{d\xi}{dz} = \frac{2\pi \alpha' \pi_\xi z^2 \sqrt{1-v^2}}{\sqrt{R^4-(2\pi \alpha'\pi_\xi)^2z^4}}. \label{eomconformal}
\end{equation}

The equation above differs from the finite temperature case since there is no blackening factor
the numerator inside the square-root is always positive,  the isolated quark is  a time-like particle, thus
we have always $v^2<1$. But we still need $\xi(z)$ to be real valuated if we want the solution to represent classical string configurations, which requires that the denominator inside the square-root should be always positive:  $R^4-(2\pi \alpha'\pi_\xi)^2z^4 >0 $ everywhere in the bulk. In this case $z\in (0,\infty)$, the reality condition for $\xi'$ requires that $\pi_\xi\leq\min{\frac{R^2}{2\pi \alpha' z^2}}$, which constrains the conserved momentum to vanish: $\pi_\xi=0$. Consequently, from eq.(\ref{eomconformal}) we find that $\xi(z)=0$ implying that all of the pieces of the string move in parallel with the same velocity.  

The drag force on the quark probe is given by the world-sheet momentum $\Pi^z_{~x} = \frac{\delta \mathcal{S}}{\delta\partial_z x}$. Thus,
we find  that the drag force vanishes:

\begin{equation}
    F_{Drag} = \Pi^z_{~x}=-\pi_\xi =0.
\end{equation}

The vanishing of the drag force means that an colored probe quark can move freely in the vacuum  of
$\mathcal{N}=4$ SYM. This scenario give us a holographic picture that shows us how the probe quark in this sort of background is not confined. Summarizing, confinement and drag force, on the holographic perspective, are connected, i.e., one is a direct manifestation of the other. This non-confining assertion observed in $\mathcal{N}=4$ SYM is extended to other non-confining gauge theories with a holographic dual. This statement will be deeply explored in the next sections.

\section{The vacuum of the Soft-Wall model}\label{section-3}
In this section we consider the soft-wall AdS/QCD model \cite{Karch:2006pv} with static quadratic dilaton defined as $\Phi(z)=k^2z^2$. The presence of a bulk dilaton field  breaks  conformal symmetry and generates confinement in the  boundary gauge theory \cite{Andreev:2006ct}. Here we will take a probe quark initially moving in the vacuum of the boundary gauge theory. The probe quark is represented in the soft-wall model bulk  by an open string with one endpoint attached to the quark location at the boundary and the other endpoint going down to $z\to \infty$. The proposed dual picture for the zero temperature case is just an extension to the original proposal  relating a heavy probe quark moving in a hot media and the  open string in   $AdS_5$ space done in \cite{Gubser:2006bz} and applied to the soft-wall model at finite temperature in \cite{Nakano:2006js}. 

The bulk geometry  is given in Poincar\'e coordinates by

\begin{equation}
 dS^2 = R^2\frac{e^{\Phi(z)}}{z^2}\,\left[-dt^2+d\vec{x}^2+dz^2\right]. \label{swmetric}
\end{equation}

In order to find the world-sheet configuration we employ the usual strategy and adopt the  ansatz: 
$t=\tau, z=\sigma,~~X^\mu(t,z) = (t, v t + \xi(z),0, 0, z)$. For such a configuration, the Nambu-Goto action is

\begin{equation}
 \mathcal{S}_{sw} = \frac{R^2}{2\pi \alpha'}\,\int dt\,dz\,\frac{e^{\Phi}}{z^2}\,\sqrt{1-v^2 +\left(\frac{d\xi}{dz}\right)^2}. \label{swaction}
\end{equation}

The Lagrangian density  is again a function of $z$ and $\xi'$ only. As in the conformal case, we have the conjugate  momentum  conservation   for $\xi$ coordinate as $\pi_\xi = \frac{\partial\mathcal{L}_{sw}}{\partial \xi'}$, leading to 

\begin{equation}
\pi_\xi = \frac{R^2}{2\pi\alpha'}\frac{\xi'e^\Phi}{z^2\sqrt{1-v^2 + \xi'^2}}.
\end{equation}

The dilaton appears explicitly in the momentum conservation expression, thus in the  equation of motion it also makes itself manifest:

\begin{equation}
 \frac{d\xi}{dz} = 2\,\pi\,\alpha'\, \pi_\xi\, z^2\sqrt{\frac{1-v^2}{R^4\,e^{2\Phi} - (2\pi\,\alpha'\,\pi_\xi)^2 z^4}}.
\end{equation}

The above expression is equivalent to the Euler-Lagrange equation that follows from the action given in eq.(\ref{swaction}).
The presence of the dilaton plays a non-trivial role here, since the reality condition for $\xi$ in this case requires that
$R^4e^{2k^2z^2} -  (2\pi\alpha'\pi_\xi)^2 z^4>0$. This condition does not require $\pi_\xi$ to vanish. It just imposes that $ \pi_\xi \leq \text{min}\left(\frac{R^2e^{k^2z^2}}{2\pi\alpha' z^2}\right)$, implying the only restriction we have is 

\begin{equation}
 \pi_\xi \leq \frac{R^2ek^2}{2\pi\alpha'}.
\end{equation}

The conserved momentum is not constrained to vanish. It has a positive upper bound and is allowed to assume any value in the range $[0, \frac{R^2ek^2}{2\pi\alpha'}] $.  Unfortunately, the condition that $\xi(z)$ is real everywhere does not determine the conserved momentum as a function of the velocity, as it happens at the finite temperature case. All we have at hand is the range of admissible classical values for the conserved momentum. The existence of this conserved momentum along the holographic coordinate in the bulk is associated with a drag force on the moving probe quark at the boundary. The rate of momentum loss due to the drag force is  

\begin{equation}
 \frac{dp}{dt} = -\Pi^z_{~x} = -\frac{\delta S_{sw}}{\delta (\partial_z x)}=  -\pi_\xi,
\end{equation}

Consequently, for any non-vanishing momentum $\pi_\xi$,  the quark moving on the vacuum of the boundary theory will be subjected to a drag force. This force acting on the moving probe quark is due to the vacuum itself. This result tells us that, in fact, \emph{a confining gauge theory has a mechanism that prevents a single quark moves freely on the vacuum}. If the probe quark is thrown into a vacuum,  this acts as a viscous media and exerts a drag force on the quark. 

 The existence of a drag force at the zero temperature in the context of the soft wall AdS/QCD model is a non-trivial phenomenology associated with the breaking of the conformal symmetry. 
 The dilaton in the soft wall model is responsible for guaranteeing the confinement of quarks. In the case of a $q\bar{q}$ pair, it makes itself manifest by providing a linear term in the interaction potential \cite{Andreev:2006ct}. But in the present case, it manifests by the presence of a drag force that acts on a quark moving in the vacuum. In this sense,  we can expect that in any confining gauge theory an isolated quark moving on the vacuum will be subjected to a drag force, interpreted as the manifestation of confinement. It is important to remark that the quark is assumed to move at a constant velocity and it is infinitely massive. The effects on the energy loss of an isolated quark due to its accelerated motion are discussed in \cite{Chernicoff:2008sa, Chernicoff:2009ff, Caceres:2010rm}. The presence of a drag force in the case of constant velocity represents a different mechanism for energy loss of the heavy quark in addition to the radiation emission due to the acceleration.
 
 \section{The general case of a gauge theory with holographic dual}\label{section-4}
 
 The usual approach to discuss confinement is to consider a static quark/anti-quark configuration and to compute the expected value of the Wilson loop operator. The confinement comes from the area law in the Wilson loop operator or equivalently in the linear behavior of the interaction energy as a function of the quark pair separation. In this sense, the seminal work of Sonnenschein \emph{et. al.}  \cite{Kinar:1998vq} classify a gauge theory in confining or non-confining according to the geometric structure of its holographic higher dimensional dual description. Here we use an alternative approach to discuss confinement by probing the vacuum of the gauge theory with an isolated quark and looking for a physical mechanism that prevents its free motion.
 
In a non-confining $\mathcal{N}=4$ SYM theory, an isolated quark can move on freely on the vacuum while in the confining theory described by the Soft-Wall model an isolated quark moving on the vacuum is subjected to a drag force. These two examples point out a possible relation between confinement and drag force. From this perspective, it seems that the drag force is a consequence of confinement.  In this section, we will use the nice results of \cite{Kinar:1998vq} in order to show that for a large class of gauge theories with holographic dual, an isolated quark moving on the vacuum of such gauge theories will be subjected to a drag force if and only if the gauge theory is confining.

In a curved background, the most general metric reads 

\begin{equation}
 ds^2 = - G_{00}(z)dt^2 +  G_{x_{T}x_{T}}(z)dx_{T}^2 + G_{zz}(z)dz^2 .
\end{equation}

We follow the proposal by Gubser that the drag force on a  massive quark, moving at the boundary, is proportional to the momentum along the holographic
direction in the dual worldsheet. This worldsheet is parametrized as $x^\mu(t,z) = (t,x_{||}(t,z), 0, 0, z)$ where we fix $\tau = t,~\sigma = z$.
The shape of the world sheet is assumed to be of the form $x_{||}(t,z) = v t + \xi(z)$.
The Nambu-Goto action for this world-sheet configuration is

\begin{multline}
    \mathcal{S}_{NG} = \frac{1}{2\pi\alpha'}\int dt\, dz\,\left[
     -G_{00}G_{zz} -  G_{00}G_{xx}\left(\frac{d\xi}{dz}^2\right) \right.\\ \left.-G_{xx}G_{zz}v^2\right]^{1/2}  
\end{multline}

The Lagrangian density is $\xi$--independent, therefore we have associated a conserved momentum  $\pi_\xi = \frac{\delta S_{NG}}{\delta \xi'(s,t)}$ allowing for a first order equation of motion with a form

\begin{equation}
 \frac{d\xi}{dz} = \frac{\pi_\xi}{2\pi\alpha'}\sqrt{ \frac{G_{00}G_{zz}  - G_{xx}G_{zz}v^2}{G_{00}^2G_{xx}^2 - G_{00}G_{xx} \pi_\xi^2} }.
\end{equation}

Due to the form of the above equation of motion, we noted that it is not necessary to assume the bulk metric as diagonal. We restrict our discussion to the cases where the bulk geometry admits a Poincar\'e-like parametrization with  Poincar\'e invariant  4-dimensional slices. 
This represents most of the AdS/QCD backgrounds at zero temperature. We assume that bulk metric has the following  form

\begin{equation}
ds^2=e^{A(z)}(-dt^2+d\vec{x}^2) + e^{B(z)}dz^2.    \label{metric1}
\end{equation}

The metric components of the bulk geometry are non-vanishing smooth functions, that near the boundary $z\to 0$ satisfy $A,B\sim -2\ln z$. The general structure of the metric in eq.(\ref{metric1}) represents the zero temperature versions of the bulk metrics  that  mimic the QCD equation of state at zero chemical potential \cite{Gubser:2008ny}. For these geometries, we find

\begin{equation}
 \frac{d\xi}{ds} =  2\,\pi\,\alpha'\pi_\xi\,e^{\frac{B(z)-A(z)}{2}}  \sqrt{ \frac{ 1 - v^2}{e^{2A(z)} - (2\,\pi\,\alpha'\pi_\xi)^2} } \label{bulkeom}
\end{equation}

In order to $\xi(z)$, and consequently $x(t,z)$,  be real valuated it is necessary and sufficient that $e^{2A(z)} -  (2\,\pi\,\alpha'\pi_\xi)^2> 0$. The conserved momentum  $\pi_\xi$ is constrained by \textcolor{black}{ $\pi_\xi < \frac{1}{2\,\pi\,\alpha'}\,e^{A(z)}$. The constraint on $\pi_\xi$  is satisfied if and only if $0\leq \pi_\xi\leq \text{min}\left(\frac{e^{A(z)}}{2\,\pi\,\alpha'}\right)$.} Hence, it will be allowed a non-vanishing $\pi_\xi$ whenever the function $f(z) = e^{A(z)}$ have a non-vanishing minimum at some $z_0$, say $f'(z_0)=0,~f(z_0)>0$. In the case where $\pi_\xi\neq 0$, the quark attached at the string endpoint is subjected to  drag force
$F_{Drag} = -\pi_\xi$. Thus we find in this form that if $f(z)$ has a non-vanishing minimum then a moving quark in the vacuum of the dual gauge theory will be subjected to a drag force. We also note that, for the classes of bulk geometries we consider, the existence of a non-vanishing minimum for $f(z)$ corresponds precisely to the same requirement that the bulk geometry should satisfy if the dual gauge theory is confining in the infrared, as it was proved in \cite{Kinar:1998vq}. Therefore, we prove that in any confining gauge theory with a holographic dual, whose bulk geometry has a form eq.(\ref{metric1}), there will be a drag force experienced on any isolated quark that moves in the vacuum. 
On the other hand,  for a non-confining gauge theory with a holographic description admitting a bulk metric with the form of  eq.(\ref{metric1}), we have (following \cite{Kinar:1998vq}) that either $f(z)$ is not bounded from below or if it has a minimum at some $z_0$ where it vanishes, i.e.,  $f(z_0)=0$. In both cases, the constraint from eq.(\ref{bulkeom}) requires that the momentum along the string vanishes, $\pi_\xi=0$.
Consequently, there is no drag force on the moving quark. This feature is realized in the particular case of the vacuum of $\mathcal{N}=4$ SYM, discussed in Section \label{section-2}, and is generalized here for the non-confining gauge theories with bulk geometry in the form of eq.(\ref{metric1}).

The physical background of the relation proved here in a holographic approach is very nice. If the gauge theory is confining then an isolated quark, initially moving with constant velocity, fells a force due to the vacuum, and hence the isolated quark cannot exists as a free particle: no forces act in a free particle. However, if there is no confinement in the gauge theory the isolated moving quark with constant velocity will not experience absolutely any resistance for its motion and it will move as a free particle. Therefore, it seems that in a confining gauge theory the concepts of fundamental particle and free particle do not mix up. 

The analysis of the single quark moving across the vacuum is not usual in the literature. But some attention to this issue has been paid in ref.\cite{Talavera:2006tj} in the context of the Sakai-Sugimoto model. There the drag force is interpreted as an instability of the system composed of an isolated quark in the vacuum of such a confining gauge theory.   We also remark that the existence of a drag force due to the interaction of the quark with the vacuum was expected in previous holographic analysis to be ruled out, such as in ref. \cite{AliAkbari:2011ue}. 

 In section \ref{section-3} we have analyzed the soft wall  model \cite{Karch:2006pv}, that is a bottom-up AdS/QCD model. In the next section, we will address top-down models, specifically D3/D7 and Sakai-Sugimoto, that are realizations of the AdS/CFT correspondence that have been proven to be confining by the area law of the $q\bar{q}$ Wilson loop operator \cite{Brandhuber:1998er}.

\section{Drag force in the D3/D7 system}\label{section-5}
In the D3/D7 system, we consider the limit $N_c \gg N_f$,  where a the set of $N_f$ D7-branes act as \emph{probes}, implying that the backreaction of them can be neglected \cite{Karch:2002sh}. The confinement arises from the embedding of the D7 probes into the D3-background.  Quarks in this scheme appear as the Chan-Patton factors of strings hanging from the D3 stack to the D7 probe (3-7 string). The conformal boundary appears in the radial (perpendicular to the D3-branes) direction of this configuration, thus the conformal boundary will localize at $r\to \infty$. 
 
 Therefore, our trailing heavy quark can be considered as the D3-brane end of the 3-7 string moving at the conformal boundary. Let us describe the geometry for this configuration. 

 The starting point is the general Dp-brane metric at $T=0$,  given by the 10-dimensional SUGRA solution
 
  \begin{equation}
dS^2=H^{-1/2}\,\eta_{\mu\,\nu}\,dx^\mu\,dx^\nu+H^{1/2}\left[dr^2+L^2\,d\Omega_{8-p}^2\right],     
 \end{equation}
 
  \noindent with 
  
   \begin{equation}
H(r)=\left(\frac{L}{r} \right)^{7-p}.   
 \end{equation}
 
 \noindent with  $L$ the gravitational length scale that controls the non-perturbative limit in the gauge theory. This scale is defined in terms of the string coupling $g_s$ and the string length $l_s$ as

\begin{equation}
L^{7-p}=\frac{g_s\,N_c}{4\,\pi}\left(4\,\pi\,l_s^2\right)^{\frac{7-p}{2}}\Gamma\left(\frac{7-p}{2}\right). 
\end{equation}

In the case of $p=3$, $L$ defines the AdS$_5$ radius.

The Dq-probe branes are spanned in the $\{0,1,2,3, \ldots, ...\,p,\ldots, q\}$ directions of the 10-dimensional flat target space. The set of D3-branes is \emph{embedded} into the D7-branes, according to the coordinates $\nu \in \{0,1,2,3, \ldots, ...\,p\}$. The brane intersection defines a $\mathcal{N}=2$ supersymmetric field theory with fundamental fermions (quarks) living in $d+1$-dimensional conformal boundary. This intersecting geometry is described by a \emph{wrapped} $S_n$ sphere inside the $S_{8-p}$ sphere, where $q=d+n+1$, with $d$ the euclidean directions along the D3-system. In our particular case, considering $d=3$ and $p=4$ leave us with $n=3$ as the dimension of the wrapping sphere and $q=7$. This embedding process defines confinement.

In the case of the  D3-brane set embedded into the D7-brane probe system, we can write the metric as 

\begin{equation}
dS^2=\left(\frac{L}{z}\right)^2\,\eta_{\,\mu \nu}\,dx^\mu\,dx^\nu+\left(\frac{z}{L}\right)^2\left[\frac{L^4}{z^4}\,dz^2+L^2\,d\Omega_5^2\right],
\end{equation}

\noindent where we have done the transformation $r=L^2/z$. Now the boundary lies at $z\to0$. In this chart, the embedding of the D3 into the D7 can be parametrized by a \emph{azimuthal} function $\psi=\psi(z)\equiv\cos \theta$, where  $\theta$ is the azimuthal angle in the $S_5$ sphere. This function controls the perturbations and will give rise to the meson masses spectrum in the flavor representation of the D7 \cite{Erdmenger:2007cm}. With this choice, we write the 5-sphere metric as

\begin{equation}
d\Omega_5^2=\frac{\psi'^2}{1-\psi^2}\frac{L^4}{z^4}\,dz^2+\left(1-\psi^2\right)^2\,d\Omega_3^2+\psi^2\,d\Omega_1^2,     
\end{equation}

\noindent where the prime denotes the  $z$ coordinated derivative, as we did before. Since we are considering free quarks moving on the D3-branes, we can keep just with the $z$-part of the $S_5$ sphere. Therefore, the D7 embedded geometry is given by

\begin{multline}
    d^2_\text{D7}= \left(\frac{L}{z}\right)^2\,\eta_{\,\mu \nu}\,dx^\mu\,dx^\nu\\+\left(\frac{L}{z}\right)^2\left[\frac{1-\psi^2+L^2\,\psi'^2}{1-\psi^2}\right] \,dz^2.  
\end{multline}

The quark moving on the $3+1$-dimensional boundary can be parametrized by the 3-7 string in the static gauge in the D3-coordinates as $x_1(t,z)=v\,t+\xi(z)$ with $x_2=x_3=0$, and $\xi(z)$ defines the profile of the 3-7 string. Following the same procedure as we did above with the other geometries, we can construct the worldsheet metric, parametrized as $\sigma_\alpha\in\{t,z\}$, as follows

\begin{equation*}
    dS^2_\text{D7}=dS^2_\text{String}=h_{\alpha\beta}\,d\sigma^\alpha\,d\sigma^\beta
\end{equation*}

\noindent with

\begin{multline}
  dS^2_\text{String}=  \left(\frac{L}{z}\right)^2\left[-(1-v^2)\,dt^2+2\,v\,\xi'\,dt\,dz\right]\\
  +\left(\frac{L}{z}\right)^2\left[1+\xi'^2+\frac{L^2\,\psi'^2}{1-\psi^2}\right]dz^2.
\end{multline}

The Nambu-Goto action in this case is 

\begin{multline}
I_\text{NG}=   -\frac{\tau}{2\,\pi\,\alpha'}\int_0^{z_\text{min}}{dz\,\left(\frac{L}{z}\right)^2\,\left[1-v^2+\xi'^2\right.}\\
\left.+\frac{L^2\,\psi'^2(1-v^2)}{1-\psi^2}\right]^{1/2},     
\end{multline}

\noindent where $z_\text{min}$ is defined by the quark mass in the model, when $\psi(z)$ is minimized, i.e., if  $\psi'(z_\text{min})=0$ implies $z_\text{min}=\frac{1}{M_q}$  \cite{Babington:2003vm,Erdmenger:2007cm}. This particular point corresponds with the point where the $S_5$ shrinks. The drag force along the $x_1$-direction is defined as

 \begin{equation}
 F_\text{Drag}=\Pi^z_{x_1},	
 \end{equation}
 
  \noindent where $\Pi^z_x$ is the canonical momentum calculated from the NG action as
  
\begin{equation}
	\Pi^\alpha_\mu
=\frac{\delta\,I_\text{NG}}{\delta\left(\partial_\alpha\,x^\mu\right)}=-\frac{1}{2\,\pi\,\alpha'}\,\sqrt{-h}\,g_{\mu\nu}\,\partial_\beta\,x^{\mu}\,h^{\alpha \beta}.
\end{equation}

In our case at hand, 

\begin{equation}
 2\,\pi\,\alpha'\,\Pi_{x_1}^z=\frac{L^2}{z^2}\frac{\xi'}{\sqrt{\xi'^2+\left(1-v^2\right)\left(\frac{1-\psi^2+L^2\,\psi'^2}{1-\psi^2}\right)}}.   
\end{equation}

From this relation we obtain the expression for $\xi'$ as

\begin{equation}
\xi'=\pm\frac{\sqrt{\left(1-v^2\right)\left(\frac{1-\psi^2+L^2\,\psi'^2}{1-\psi^2}\right)}}{\sqrt{1-\frac{z^4}{L^4}\,\left(2\,\pi\,\alpha'\,\Pi_{x_1}^z\right)^2}}\,\frac{z^2}{L^2}\,\left(2\,\pi\,\alpha'\,\Pi_{x_1}^z\right).    
\end{equation}

In Poincar\'e coordinates it means that the D7 probe brane 
extends over the range $z\in (0,z_\text{min}]$. At $z= z_\text{min}$ the $S_5$ sphere shrinks and at this point the trails string ends. The $z$ coordinate is bounded from above and it leads to the range of allowed conserved momentum:  $0\leq \Pi_{x_1}^z \leq \frac{M_q^2}{2 \pi \alpha'}$. In this case where the energy scale $z_\text{min}$ is  responsible for the  confinement at the dual gauge theory in a top-down frame,  we also find the presence of a drag force of an isolated quark due to the non-perturbative vacuum of the gauge theory. 

\section{Drag force in the Sakai-Sugimoto model}\label{section-6}
We consider in this section an isolated quark moving on the vacuum of the gauge theory dual to the $D4/D8$ background of the  holographic model proposed by Sakai and Sugimoto \cite{Sakai:2004cn, Gao:2006uf}. As an application of the general result given in section \ref{section-4}, the confining bulk geometry will be responsible for the presence of a drag force experienced by the moving quark  in the vacuum of the dual gauge theory. The zero temperature geometry   of the gravitational background  is given by

\begin{multline}
   ds^2 = \left(\frac{U}{R}\right)^{\frac{3}{2}}(\eta_{\mu\nu}dx^\mu dx^\nu +f(U)d\tau^2) \\
   + 
    \left(\frac{R}{U}\right)^{\frac{3}{2}}\left(\frac{dU^2}{f(U)} + U^2d\Omega_4^2\right), 
\end{multline}

\noindent where $x^\mu$ are the coordinates on the   4-dimensional boundary theory. The warp function $f(U)$  is given by
 
 \begin{equation}
     f(U) = 1- \frac{U_{KK}^3}{U^3}.
 \end{equation}
 
 Notice that the radial coordinate $U$ is bounded from below $U\geq U_{KK}$ and the coordinate $\tau$
 is periodic with period $\tau_0 = (4\pi/3)\sqrt{R^3/U_{KK}}$.  
 
 In this scenario, the dual picture of a free quark moving in the vacuum of the dual gauge theory is a classical string attached to the quark stretching along the radial $U$ direction in the bulk. The Gubser ansatz for the string parametrization gives $x^\mu(t,U) = (t, x(t,U),0,0,U)$, where $x(t,U) = vt + \xi(u) $ and the compact directions in the bulk are keep constant. The Nambu-Goto action for this string configuration is given by

 \begin{equation}
    S_{NG}= \frac{1}{2\,\pi \,\alpha'}\int dt\,dU \sqrt{\frac{1-v^2}{f(U)}+ \frac{U^3}{R^3}\left(\frac{d\xi}{dU}\right)^2}. 
 \end{equation}
 
  The conservation of  $\pi_\xi=\delta S_{NG}/\delta\xi'(t,U)$ results in the first order differential equation
 
 \begin{equation}
     \frac{d\xi}{dU} =  2\,\pi\,\alpha'\, \pi_\xi  \left(\frac{R}{U}\right)^{\frac{3}{2}}\sqrt{\frac{1-v^2}{\frac{U^3}{R^3}-
    ( 2\,\pi\,\alpha'\ \pi_\xi)^2} }.
 \end{equation}
 
The radial direction is bounded from below as $U\geq U_{KK}$ where the trailing string ends. The above expression will be real valuated provided that $\pi_\xi\leq  \frac{1}{2\,\pi\,\alpha'} \left(\frac{U_{KK}}{R}\right)^{\frac{3}{2}}$. As it was argued in \cite{McNees:2008km} the limit $U_{KK}\to 0$
 corresponds to the supersymmetric case where there is no confinement. On the other hand, we find here that there is a drag force in the vacuum of the dual gauge theory if  $U_{KK}>0$. This means that there will be a drag force at zero temperature even that the gauge theory stays in the confining region. This case represents another example of a top-down model that realizes a direct relation between confinement and drag force at zero temperature.
 
 \section{Estimating the Fraction of Energy Loss}\label{section-7}
 We have shown that at zero temperature an isolated probe quark fells the effect of a drag force when it is moving in the vacuum of a confining gauge theory. Unfortunately, the usual constraints of classical mechanics do not allow for a complete determination of the magnitude of the drag force as a function of the velocity of the quark. Here we will use some physical arguments to estimate this functional dependence and the fraction of energy transferred to the vacuum by the moving quark.
 
 A naive way to fix $\pi_\xi(v)$ at zero temperature is by considering the finite temperature expression, and then calculating the limit where $T\to0$. We take, for example, the case of the soft wall model \cite{Nakano:2006js}, that at finite temperature   it is found that 
 
  \begin{equation}
       \pi_{sw}(v,~T) =  \frac{\pi^2 \,T^2 \,R^2}{2\,\alpha'}\,e^{\frac{k^2\sqrt{1-v^2}}{2\pi^2 T^2}}\, \frac{v}{\sqrt{1-v^2}}.
 \end{equation}
 
 The corresponding zero temperature limit  is divergent, i.e.,  $T\to 0\Rightarrow \pi_{sw}\to\infty$. The expression for $\pi_{sw}(v,~T)$ comes from the presence of a black hole in the bulk geometry. But this particular setup is not stable at low temperatures. At some critical temperature $T_c$, the bulk geometry undergoes a Hawking-Page transition from the black hole phase to a thermal AdS phase, where there is no black hole. The critical temperature for the  Hawking-Page transition \cite{Hawking:1982dh} is zero for conformal gauge theories and is non-vanishing for non-conformal gauge theories \cite{Witten:1998zw,Herzog:2006ra}. In a confining gauge theory, we cannot obtain the zero temperature expression for $\pi_\xi$ by performing the $T\to0$ limit in the expression obtained from the black hole phase of the bulk geometry.
 
 We assume that if the quark does not move, there is no drag force acting on it, i.e., $\pi_\xi(0) = 0$. The drag force comes from the relative motion between the quark and an external observer that looks at the vacuum state of the gauge theory. For very small velocities,  the drag force should be linear in the velocity ($v\ll 1 \Rightarrow \pi_\xi \sim v$). On the other hand, special relativity imposes that $v< 1$ for a time-like particle, such that the maximum drag will be reached when $v\to 1$ or, equivalently, in the limit $\gamma v \to \infty$, where $\gamma = 1/\sqrt{1-v^2}$ is the Lorentz factor. Thus, we  use here  an estimation of $\pi_\xi(v)$  given by the following smooth function interpolating both regions, i.e.,
 
 \begin{equation}
 \pi_\xi(v) = \pi_\text{max}(1-e^{-\gamma v}).
\end{equation}

In the above expression $\pi_\text{max} =  \text{min}\left(\frac{e^{A(z)}}{2\,\pi\,\alpha'}\right)$, where $A(z$) is defined in eq. \eqref{metric1}. For each model considered here, we have the following expressions for the transferred momentum

\begin{eqnarray}
\pi_\text{SW}&=&\frac{R^2}{2\,\pi\,\alpha'}\,e\,k^2,\\
\pi_\text{SS}&=&\frac{1}{2\,\pi\,\alpha'}\left(\frac{U_{KK}}{R}\right)^{3/2},\\
\pi_\text{D3/D7}&=&\frac{1}{2\,\pi\,\alpha'}\,M_q^2\,L^2.
\end{eqnarray}

The large mass of the heavy quarks allows us to model their classical motion as
non-relativistic, since most of its energy  will be inertial. Therefore we can take the approximation $\gamma v \ll 1$. In this   approximation we have that $e^{-\gamma v} \simeq 1-\gamma v$ leading to 
a momentum $\pi_\xi = \pi_{max}\gamma v$. The holographic dictionary maps this conserved momentum along the holographic coordinate to the momentum loss by the dual heavy quark on the boundary
$\frac{dp}{dt} = -\frac{1}{2\pi \alpha '}\pi_\xi$. Thus, we have  that

\begin{equation}
 \frac{dp}{dt} = - \frac{\pi_{max}}{2\,\pi \,\alpha '} \gamma v = -\frac{\pi_{max}}{2\,\pi\,\alpha'm_q}p(t),
\end{equation}

\noindent where the relativistic momentum is $p= m_q  \gamma v$. The above equation is a first-order one and can be directly  integrated, obtaining the loss of momentum by the heavy quark moving on the vacuum

\begin{equation}
    p(t) = p(0) \,e^{-\frac{t}{t_0}},~~ t_0 = \frac{2\,\pi\,\alpha'\,m_q}{\pi_{max}}. \label{plost}
\end{equation} 

Notice that  in the above equation  we have defined the characteristic time $t_0$, which in the finite temperature case, is interpreted as a diffusion time. Nevertheless, in the present case, where the medium of propagation is the vacuum, a different interpretation is required.

Usually, the interpretation is related to the jet formation phenomena as the vacuum analogue of thermal diffusion.  We also notice that for an isolated quark $q$ the time is $t_0\sim 1/\pi_{max}$ with $\pi_{\max} = f(z_0)$, while for the corresponding $q\bar{q}$ meson, the string tension  (obtained from Wilson Loop calculations) is given by $\sigma = f(z_0)$. Hence, our estimation of $\pi_\xi(v)$ reveals that the diffusion time is inversely proportional to the string tension, i.e., $t_0\sim \frac{1}{\sigma}$. The stronger the $q\bar{q}$ pair is glued together the fast one of them ($q$ or $\bar{q}$) will hadronize in the vacuum.

We use the estimation for the time dependence of the quark spatial momentum, eqn.\eqref{plost}, to estimate the fraction of the initial quark energy transferred to the medium (the total energy of the formed jet). The energy of a particle with spatial momentum $p$ and rest mass $m$ is given by $E(p) = \sqrt{m^2+p^2}$. We suppose that the quark lost energy is until $t=t_0$ and let $p(0)=p_0$ be the initial (production) quark momentum. Then the fraction of energy lost is estimated 

\begin{equation}
    \frac{\delta E}{E} \sim \frac{E(t_0)-E(0)}{E(0)}  = -1+\frac{1}{e}\sqrt{\frac{e^2 + (\frac{p_0}{m})^2}{1+(\frac{p_0}{m})^2}}.
\end{equation}

It is remarkable that in the very heavy quark mass limit we have $\frac{p_0}{m}\to 0$, then $\frac{\delta E}{E}\to 0$. While in the limit of very light quarks $\frac{p_0}{m}\to \infty$, and hence $\frac{\delta E}{E}\to -(e^{-1}-1)\sim 0.632$. These simple estimations provide a piece of evidence that heavy quarks will transfer a smaller fraction of its initial energy to the hadronic jet formed in comparison to the energy a light quark transferred to its corresponding hadronic jet.

This phenomenon  is known as the \emph{dead cone effect} \cite{Thomas:2004ie,Xiang:2005ce,Maltoni:2016ays,Tripathee:2017ybi,Cunqueiro:2018jbh,Kopeliovich:2019jsr,Kopeliovich:2019nel}.  This  estimation suggests that the dead cone effect is a consequence of the momentum damping and it will be observed given that the final momentum is smaller the the initial momentum. If $p_f  =\frac{p_0}{\Lambda},~\Lambda>1$ we get that $\frac{\delta E}{E}\to 0$ in the limit $\frac{p_0}{m}\to 0$ while that in the limit  $\frac{p_0}{m}\to \infty$ gives  $\frac{\delta E}{E}\to -(\Lambda^{-1}-1) < 0$. In our case, the existence of a  drag force provides the momentum damping of the isolated quark that moves across the vacuum. Consequently, we expect that the dead cone effect will take place as a signature of confining in these holographic dual models.

\section{Conclusions and Final Remarks}\label{section-8}

The present results show that the property of confinement in gauge theories can be probed in several forms, making it something multifaceted. The same \emph{property} that avoids mesons to split into an isolated quark and anti-quark states also prevents them to move freely in the vacuum. However, if any of these two situations (confined or deconfined quarks) are considered to represent the initial conditions, there will be no free quarks moving on the vacuum as the final (asymptotic) state. 

On the other hand, it has been shown that for a string representing a meson, with both endpoints attached at the boundary, there is no such resistance \cite{Talavera:2006tj}. In the bulk, we face that the configuration of many disconnected (infinite) pieces of the string, representing mesons moving with a small fraction of the original momentum, is energetically favored in comparison with an infinitely long string. Consequently, the initial string, attached to the isolated quark, will break into a final state of infinitely many pieces of strings attached to $q\bar{q}$ pairs. This process gives us a holographic picture of jet formation. The physical picture of jet formation that emerges in this holographic approach matches the early proposal by Field and Feynman \cite{Field:1977fa} and agree qualitatively with the idea that there is a deep connection between quark propagation in vacuum and the hadronization process \cite{Accardi:2020iqn}.

An  important experimental aspect regarding QGP formation is the  back to back jet suppression 
\cite{Collins:1981uk, Adler:2002tq, Gyulassy:2004vg}. This phenomenon works as follows: some high energy virtual photon decays into a $q\bar{q}$ pair that moves in opposite directions with high momentum.  If the pair is created in the   QGP fireball neighborhood,   one quark will go inward the QGP medium and the other will go outward. The inward quark fells the interactions of the hot medium where it thermalizes and its associated jet is quenched,  while the outgoing quark leaves the QGP and penetrates the vacuum with large momentum producing a jet of hadrons. The present results complement the holographic picture of this physical mechanism by discussing the dynamics of the isolated quark that penetrates the vacuum. In the conformal, non-confining,  scenario there's no jet at all since the outgoing quark will move freely at a constant velocity, thus it will not hadronize. Jets in the vacuum will appear at the non-conformal, confining,  scenario only.

\begin{acknowledgments}
We acknowledge the financial support of FONDECYT (Chile) under Grants No. 1180753 (A. V.) and No. 3180592 (M. A. M. C.).
\end{acknowledgments}

\bibliography{references}
\end{document}